\documentclass[lettersize,journal]{IEEEtran}
\usepackage{amsmath,amsfonts}
\usepackage{algorithmic}
\usepackage{algorithm}
\usepackage{array}
\usepackage[caption=false,font=normalsize,labelfont=sf,textfont=sf]{subfig}
\usepackage{textcomp}
\usepackage{stfloats}
\usepackage{url}
\usepackage{verbatim}
\usepackage{graphicx}
\usepackage{cite}
\usepackage{orcidlink}
\begin{document}

\title{PolycubeNet: A Dual-latent Diffusion Model  for\\Polycube-Based Hexahedral Mesh Generation}
\IEEEaftertitletext{%
\vspace{-1.5em}
\begin{center}
  \includegraphics[width=\textwidth, trim=0.21cm 0.15cm 0.01cm 0.7cm, clip]{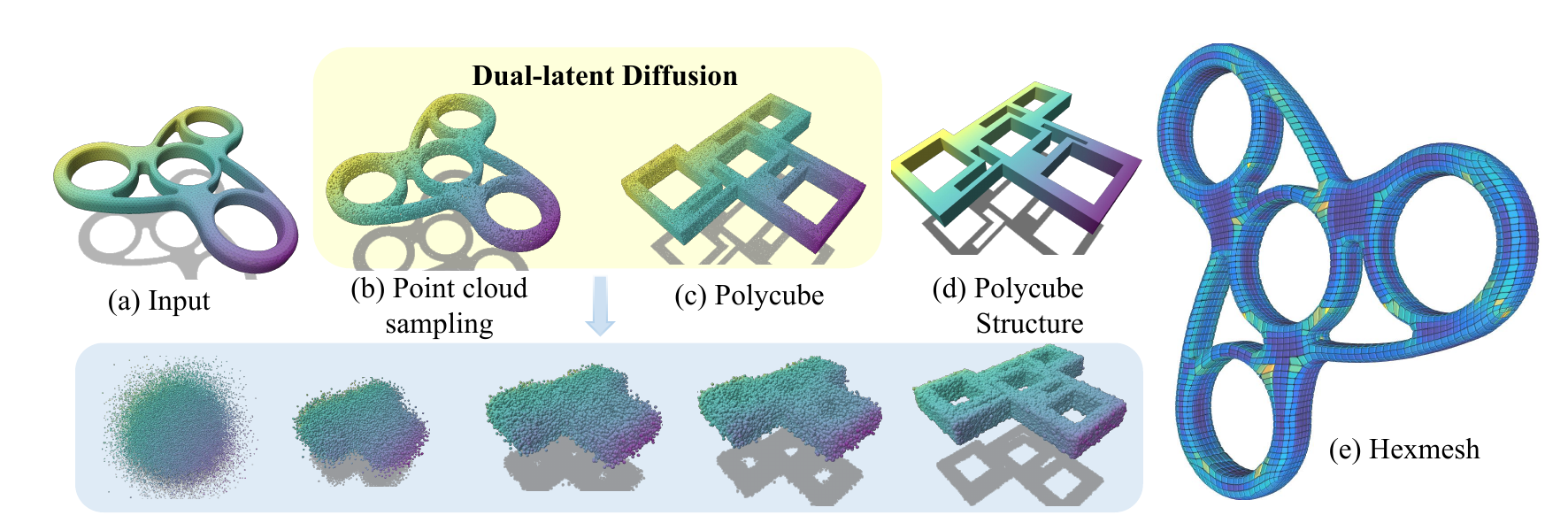}

  \vspace{0.7em}

  \refstepcounter{figure}
  \label{fig:figure1}
  \parbox{\textwidth}{\footnotesize
  \textbf{Fig.~\thefigure.} Hex-mesh generation with PolycubeNet.
  (a) Input surface mesh. (b) Poisson-disk sampled point cloud used as the condition.
  (c) Polycube point cloud generated by our dual-latent conditional diffusion model.
  (d) Recovered polycube structure after registration and structure extraction.
  (e) Hex mesh obtained by partitioning in the polycube domain and geometric pullback.
  The blue strip illustrates the reverse diffusion sampling trajectory.
  }
\end{center}
\vspace{0.4em}
}

\author{Lu He\,\orcidlink{0009-0001-6354-2558}, Qitao Deng, Junjiang Deng, Liangbin Deng, Yanjun Liang, Wenting Yang, Guoqiang Wang, and Na Lei%
\thanks{Lu He, Liangbin Deng, Yanjun Liang, and Na Lei are with Dalian University of Technology. 
E-mail: herain2008@163.com, denglb2026@163.com, 20222241293@mail.dlut.edu.cn, nalei@dlut.edu.cn.}%
\thanks{Qitao Deng is with Jiangxi University of Science and Technology. 
E-mail: 1920234411@mail.jxust.edu.cn.}%
\thanks{Junjiang Deng is with the School of Mathematical Sciences, Guangxi Minzu University. 
E-mail: 19507973898@163.com.}%
\thanks{Wenting Yang is with Caohejing Hi-Tech Park Development Co., Ltd. 
E-mail: wentingyang13@fudan.edu.cn.}%
\thanks{Guoqiang Wang is with Shanghai AI Laboratory. 
E-mail: wangguoqiang@pjlab.org.cn.}%
\thanks{Na Lei and Guoqiang Wang are the corresponding authors.}%
}



\maketitle


\begin{abstract}
Hexahedral meshes are widely used in simulation pipelines, yet automatic generation remains challenging for complex CAD geometries. Polycube-based hexahedral meshing is a representative approach due to its regular, parameterization-friendly structure, but existing polycube construction methods often rely on intricate surface segmentation and local heuristics, which can produce artifacts or fail on difficult shapes.

In this paper, we propose an end-to-end framework for polycube generation based on conditional diffusion models. Given an input geometry represented as a point cloud, our method directly produces a corresponding polycube point cloud, eliminating the need for explicit surface segmentation or predefined polycube templates. At the core of our approach is a dual-latent conditional diffusion architecture that confines computationally expensive self-attention operations to a fixed-capacity, low-dimensional latent space. This design effectively decouples computational complexity from the resolution of both the input geometry and the output polycube, thereby avoiding the quadratic cost typical of point cloud self-attention mechanisms while supporting flexible input and output resolutions.

To obtain a hexahedral mesh, the generated polycube is aligned to the input shape via rigid and non-rigid point cloud registration to establish surface correspondence, followed by a polycube-to-hex pipeline. We additionally create and release a paired dataset of CAD meshes and their corresponding polycube meshes, together with the core implementation of our model. Experiments show that PolycubeNet generalizes to complex CAD models with arbitrary genus and produces high-quality polycube structures within seconds, improving robustness and efficiency over prior learning-based approaches.
\end{abstract}

\begin{IEEEkeywords}
hexahedral mesh generation, diffusion models, generative modeling, polycube-based meshing
\end{IEEEkeywords}

\section{Introduction}
\IEEEPARstart{T}{he} quality and efficiency of mesh generation directly impact the accuracy and performance of numerical simulations, making it a critical bottleneck in integrated CAD/CAE workflows \cite{Hexmesh_survey}. The NASA CFD Vision 2030 study \cite{NASA_report} highlights that meshing and mesh adaptation remain key limiting factors in high-fidelity simulations, significantly restricting the computational capabilities and efficiency of complex engineering analyses. Hexahedral meshes, in particular, offer higher accuracy and more stable performance with fewer elements compared to tetrahedral meshes, especially in scenarios involving large deformation or highly elastic/plastic materials. However, the automatic generation of high-quality hexahedral meshes for complex geometries and diverse topologies remains an elusive challenge in the field \cite{HolyGrailHex}.

\begin{figure}
    \centering
    \includegraphics[width=1\linewidth, trim=0.00cm 1.3cm 0.00cm 1.2cm, clip]{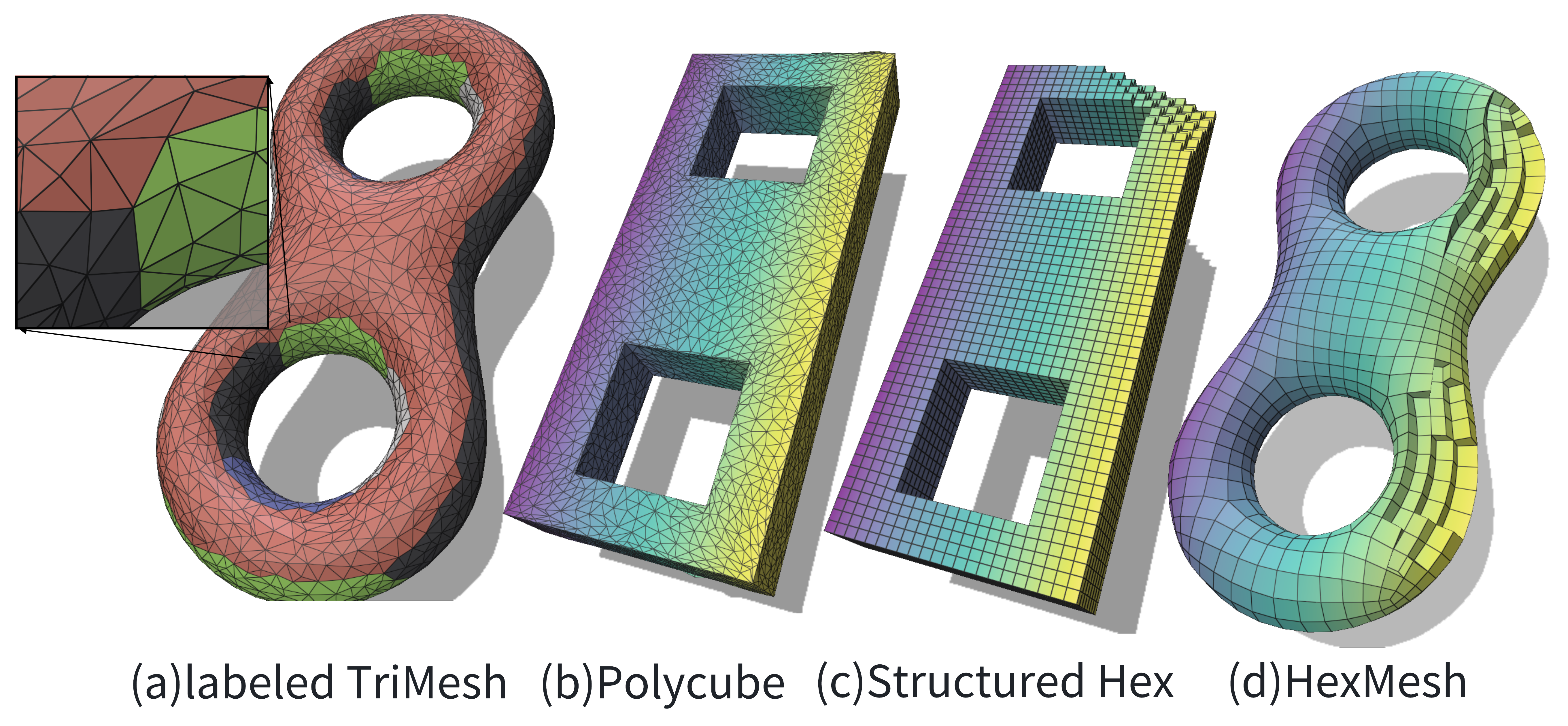}
    \caption{Classical polycube-based pipeline for hexahedral mesh generation.}
    \label{fig:tra_pipline}
\end{figure}

Polycube-based hexahedral meshing is an important class of approaches, favored for its regular structure and parameterization-friendly formulation. A polycube represents a specialized geometry whose surface elements align with one of the principal axes $(\pm X,\pm Y,\pm Z) $\cite{polycubemap04},  reducing the problem of meshing a complex shape to generating a structured mesh in a regular domain and mapping it back to the original geometry \cite{deform_polycube2011}. The classical polycube-map pipeline is summarized in Figure~\ref{fig:tra_pipline}: given an input triangular mesh $M$, it assigns each face an axial label $(\pm X,\pm Y,\pm Z)$ (typically via heuristics or local optimization), deforms/parameterizes the mesh under these constraints to obtain a polycube structure and a consistent map \cite{Polycut}, extracts a structured hexahedral mesh by integer-grid partitioning in the polycube domain, and finally pulls it back through the inverse map to obtain a hexahedral mesh on the original shape \cite{HexEx}. However, constructing a volumetric polycube domain that simultaneously satisfies axis alignment, global topological consistency, and low distortion remains challenging \cite{15fix_normal_polycube}. Existing methods still rely heavily on surface segmentation and heuristic discrete optimization, which often introduce local optima and sequential dependencies, limiting robustness on complex and high-genus geometries \cite{close_form_polycube}.

To overcome these limitations, DL-Polycube \cite{DL-Polycube} introduced a deep learning approach that classifies shapes into a limited set of polycube-structure categories, thereby reducing dependence on heuristic rules. However, because the space of possible polycube configurations is virtually infinite, this category-based strategy cannot capture the full diversity of real CAD geometries. Subsequently, DDPM-Polycube \cite{DDPM-polycube} formulates the mapping from geometry to polycube as a denoising diffusion process. Its training data are synthesized by perturbing a small set of template polycubes (primarily covering low-genus shapes, e.g., genus $0$--$2$). During inference, instead of sampling from standard Gaussian noise, it initializes the diffusion chain with the input shape’s point cloud (1024 points) to guide generation. To compensate for the resulting mismatch between the training and inference input distributions, the method further introduces a mean-compensation term $q$ as an engineering correction to the diffusion dynamics. In contrast, we formulate polycube generation as conditional diffusion: sampling always starts from Gaussian noise (at arbitrary resolution), while the input geometry is injected solely through conditioning, yielding an end-to-end generator for structurally consistent polycube point clouds without ad-hoc initialization changes. 

A practical bottleneck in diffusion over point sets is scalability: when denoising networks rely on point-token self-attention, both memory and runtime grow rapidly with point count, making high-resolution generation prohibitively expensive \cite{DDPM}. To address this, we propose PolycubeNet, a  conditional diffusion model based on a dual-latent Transformer design inspired by latent-space computation \cite{pointinfinity}. Specifically, we move the high-cost self attention computation in denoising and conditioning into a fixed-capacity latent space, where the input point cloud interacts with the latent representation via lightweight cross-attention. This design decouples the core computational complexity from the input resolution, enabling PolycubeNet to handle arbitrary-resolution inputs and outputs while filtering out redundant information and capturing global topological structure for consistent results.

Unlike classical polycube pipelines that derive polycube structures through surface labeling and heuristic optimization, PolycubeNet treats the polycube as a generative target and directly synthesizes a polycube representation conditioned on the input geometry, while keeping the subsequent parameterization and hex extraction stages unchanged. After generation, we establish stable surface correspondences via rigid alignment followed by non-rigid refinement, enabling robust polycube-to-hex construction. This structure-first strategy helps mitigate parameterization-induced structural artifacts—regular distortions and unnatural mesh organization driven by the parametric domain rather than the intrinsic geometry—as characterized in \cite{Hexmesh_survey}. 


Finally, to address the scarcity of real polycube supervision, we build and release a paired dataset of CAD geometries and polycube point clouds derived from the ABC dataset, together with the core implementation.\footnote{\textcolor{black}{Project code: \url{https://github.com/herain520/AI4polycube}; Dataset: \url{https://github.com/herain520/PolycubeDataset}.}}

The contributions of this work are as follows:
\begin{itemize}
\item \textbf{Introduction of a conditional diffusion framework for polycube generation}. Our method conditions on the input geometry to directly generate structurally consistent polycube point clouds in an end-to-end manner.

\item \textbf{A novel dual-latent Transformer architecture}. This design decouples the computational cost of the core self-attention mechanism from the point cloud resolution, supporting arbitrary input and output resolutions while preserving structural integrity.

\item \textbf{Direct prediction of polycube structure}, bypassing traditional workflows. Unlike conventional “segmentation-first” pipelines, our approach directly infers the polycube, thereby avoiding artifacts typically introduced by intermediate parameterization steps \cite{Hexmesh_survey}.

\item \textbf{Creation and public release of the first CAD-model-based polycube point cloud dataset}. This dataset addresses the critical problem of data scarcity that hinders existing learning-based methods.
\end{itemize}

\section{Related Work}

\subsection{Polycube-based Hexahedral Meshing}

Existing polycube constructions can be broadly grouped into two categories. Deformation/optimization-based methods continuously deform the surface or volume toward an axis-aligned polycube while explicitly controlling distortion and injectivity \cite{L1polycube14,L1polycube19}. Label/structure-search methods instead solve a discrete labeling problem via combinatorial optimization or search (e.g., graph cuts, perturbations, or evolutionary strategies) to obtain a globally coherent axis structure, followed by geometric realization \cite{Polycut,Evocube}. Several variants introduce richer constraints or representations, including constrained voxelization \cite{fu19polycube}, $\ell_1$-based construction \cite{L1polycube14}, intrinsic mixed-integer formulations \cite{Intrinsic_polycube}, segmentation-driven charting \cite{CVTpolycube}, and interactive decomposition for improved controllability \cite{Interactive_cuboid}.

Most pipelines further impose explicit or implicit “polycube validity” constraints to ensure the optimized result forms a usable polycube structure. In practice, commonly used constraints tend to favor $3$-connected configurations and do not fully cover the entire polycube space; \cite{expand_space} discusses expanding the solvable space via the Gauss–Bonnet theorem, yet the solver still relies on heuristic optimization and engineering repairs. Therefore, obtaining a structure-consistent polycube/parameter-domain representation that can be stably sampled and reliably pulled back—while reducing dependence on brittle heuristics and manual tuning—remains a key bottleneck. Data-driven generative modeling (e.g., conditional diffusion) provides an alternative route by directly synthesizing polycube representations consistent with the input geometry.

\subsection{AI for Meshing: Representation and Generation}
Geometric deep learning has substantially improved how triangle meshes are represented and processed, including random-walk-based sequential encodings \cite{MeshWalker}, edge-centric convolutional architectures \cite{MeshCNN}, and discretization-agnostic learning on surfaces \cite{DiffusionNet}. On the generation side, a common direction is to recover or synthesize triangle meshes from point sets, either by progressively deforming a fixed-topology template toward the target surface \cite{Point2Mesh} or by learning triangulation directly from point clouds \cite{PointTriNet}. Beyond triangle meshes, combining advancing-front procedures with supervised and reinforcement learning has also shown promise for planar quadrilateral meshing \cite{RL_quad}.
Compared to triangular or quadrilateral settings, hexahedral meshing is more constrained by discrete, highly non-local topology and strict validity/quality requirements, which makes it difficult to define stable learning targets and differentiable supervision. A practical strategy is to introduce a structured intermediate representation that separates global topological organization from geometric detail. Polycubes provide such an axis-aligned structural domain and naturally couple to classical grid sampling and pull-back. Motivated by this, we learn to generate polycubes—rather than hex elements directly—using conditional diffusion.
\begin{figure*}
    \centering
    \includegraphics[width=1\linewidth, trim=6.00cm 22cm 8.00cm 17.5cm, clip]{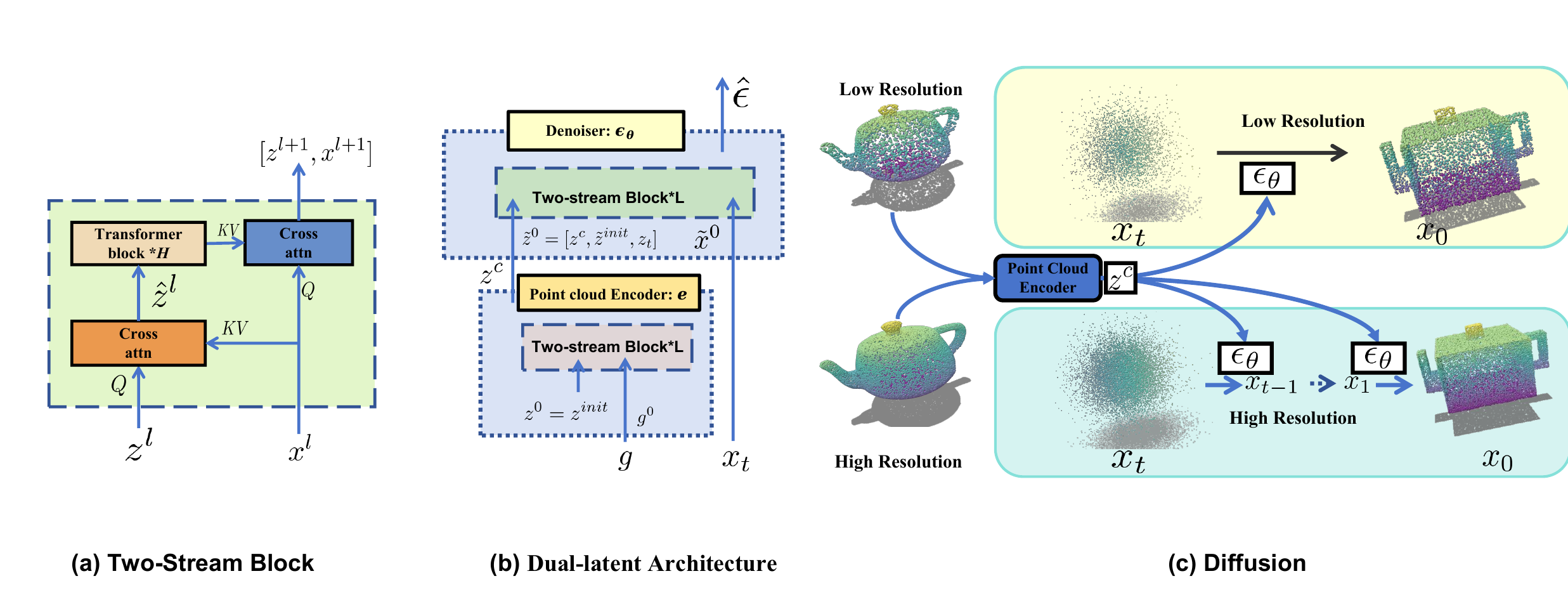}
    \caption{\textbf{Framework of PolycubeNet} for polycube generation conditioned on original geometric point clouds. 
    \text{\bf (a)} \textbf{Two-Stream block}: The two-stream contains a reading module with a across attention, a compute module with $H$ transformer blocks and a writing module with a across attention.
    \text{\bf(b)} \textbf{Architecture}: The Dual-Latent diffusion model architecture consists of two primary modules: a point cloud encoder and a denoising model $\epsilon_\theta$. Both modules utilize a two-stream transformer architecture. This specific design shifts the self-attention computations for both the conditional encoding and the denoising processes into a fixed-size latent space. \text{\bf (c)} \textbf{Core Mechanism}: During the training phase, the model operates using low-resolution representations for both the conditioning signals and the generated outputs. However, during inference, the framework is designed to support high-resolution inputs and outputs. }
    \label{fig:arch}
\end{figure*}
\subsection{DDPM}

The model in this paper is based on  DDPM \cite{ho2020denoising} which is a type of deep learning-based generative model that achieves high-quality sample generation. Its core idea is to model the data generation process as a reverse diffusion from a simple distribution (Gaussian noise) to a complex data distribution. Since the introduction of diffusion technology, it has been rapidly applied in fields such as image generation   \cite{rombach2022high, saharia2022photorealistic}, video generation \cite{blattmann2023stable, wan2025wan}, and 3D generation \cite{liu2023zero,tang2024mvdiffusion++,long2024wonder3d}.

DDPM consists of two mutually dual Markov processes: the forward diffusion process and the reverse generation process.

The forward process  adds Gaussian noise to the original data  \(\mathbf{x}_0\) at time step $t$ as:
\[
\mathbf{x}_t = \sqrt{\bar{\alpha}_t} \, \mathbf{x}_0 + \sqrt{1 - \bar{\alpha}_t} \, \boldsymbol{\epsilon}, \quad \boldsymbol{\epsilon} \sim \mathcal{N}(0, \mathbf{I}),
\]
where \(\alpha_t = 1 - \beta_t\) and \(\bar{\alpha}_t = \prod_{s=1}^t \alpha_s\) and where \(\beta_t \in (0, 1)\) is a noise scheduling parameter.

The reverse (generative) process is learned by a parameterized denoiser \(\boldsymbol{\epsilon}_\theta(\mathbf{x}_t, t)\), which predicts the noise component present in the noisy input \(\mathbf{x}_t\). Generally, it is trained by minimizing the following loss:

\[
\mathcal{L}_{2} = \mathbb{E}_{t, \mathbf{x}_0, \boldsymbol{\epsilon}} \left[ \| \boldsymbol{\epsilon} - \boldsymbol{\epsilon}_\theta( \sqrt{\bar{\alpha}_t} \, \mathbf{x}_0 + \sqrt{1 - \bar{\alpha}_t} \, \boldsymbol{\epsilon}, t ) \|^2 \right].
\]

\subsection{DDPM-polycube}

DDPM-Polycube introduces DDPM into the generation of regular polycube structures, formalizing the mapping from "arbitrary input geometry → regular polycube" as a conditional denoising process. Specifically, this method utilizes only two types of basic voxels—the genus-0 cube and the genus-1 perforated cube—combined on a 2×1 grid to produce nine candidate polycube configurations.

During the forward diffusion phase, an input point cloud of size 1024×3 is first reshaped into a pseudo-image representation of size 32×32×3. To preserve the original geometric features during the noise-adding process, the authors design a **non-standard Gaussian noise** scheduler strategy, thereby obtaining heavily distorted yet geometry-preserving initial noise samples, rather than the Gaussian white noise traditionally used in DDPM.

In the reverse generation phase, DDPM-Polycube constructs a conditional U-Net that takes the distorted geometry as noise input, progressively predicts and removes the current noise/deformation, and ultimately recovers one of the nine regular polycube point clouds. 

Although DDPM-Polycube implements the idea of generating polycube-corresponding geometry point clouds using a Denoising Diffusion Probabilistic Model (DDPM), its generated cubes are template-based, limited in variety, and geometrically simplistic. This stands in contrast to the potential complexity of the input geometry. Moreover, the U‑Net architecture adopted in the framework—which transforms point clouds into image-like representations—may result in suboptimal generation performance in practical applications. Additionally, both the input and output of the model are restricted to fixed-size point clouds containing precisely 1,024 points.
\begin{figure*}
    \centering
    \includegraphics[width=0.8\linewidth, trim=0.00cm 0.6cm 0.00cm 0.1cm, clip]{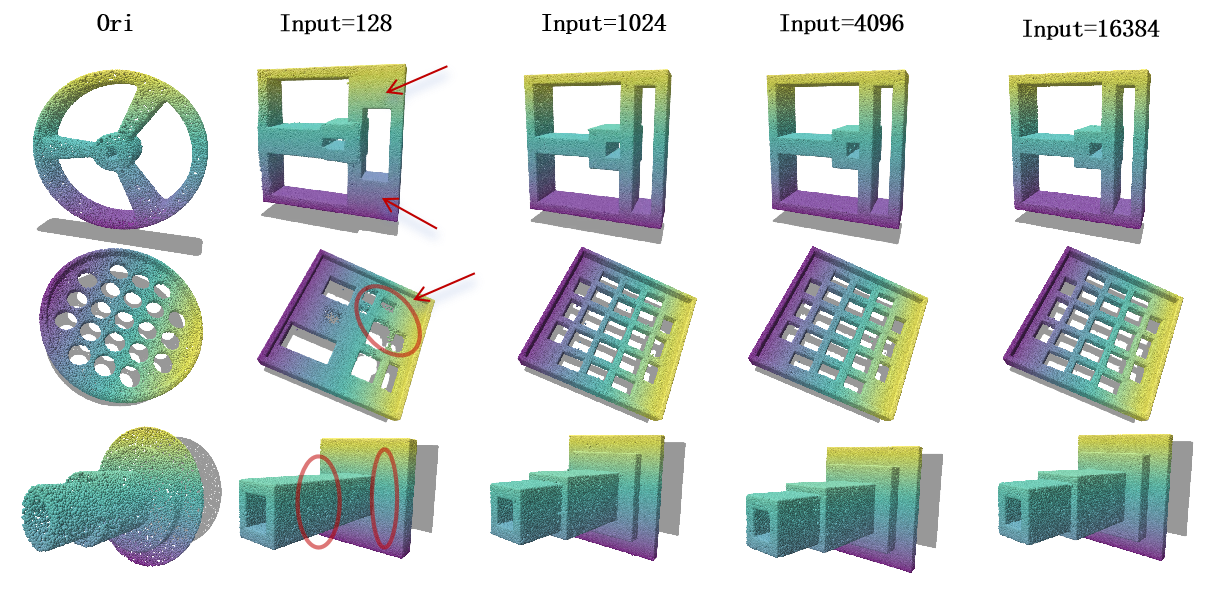}
    \caption{Results of polycube generation with incrementally increasing input resolutions from 128 to 16,384 and a fixed output resolution of 131,072.}
    \label{fig:fixed_output_resolution}
\end{figure*}

\subsection{Transformer-based DDPM for point generation}

Distinct from  PVD \cite{PVD} which combines a denoising diffusion model with a hybrid point-voxel representation, Point‑E \cite{Point-E} adopts a more concise Transformer‑based architecture and reduces the reliance on specialized 3D priors, thereby offering greater efficiency in generation. However, the Transformer framework employed in Point‑E lacks resolution invariance. Therefore, Pointinfinity introduces a two-stream transformer-based conditional diffusion model utilizing a fixed latent representation, capable of generating point clouds at arbitrary resolutions from a single image input. Specifically, the architecture models 3D shapes through a fixed-size latent stream while concurrently processing point clouds via a variable-size data stream. Information exchange between these two streams is mediated exclusively through a lightweight "read/write" cross-attention mechanism. This design reduces the computational complexity of the attention mechanism from $O(n^2)$ to being linearly proportional to the point cloud scale. Consequently, the model can be trained on low-resolution data yet remains capable of producing high-fidelity point clouds at significantly higher resolutions during the inference phase.

\section{Method}

This paper investigates the problem of automatic structured hexahedral mesh generation for geometries characterized by arbitrary topology, multiply-connected domains, high genus, and complex boundary features. Let $g \in \mathbb{R}^{N \times 3}$ represent the sampled point cloud of the original geometry, comprising 3D spatial coordinates (XYZ). Correspondingly, let $x \in \mathbb{R}^{M \times 6}$ denote the associated polycube point cloud. Each point in $x$ contains 3D spatial coordinates (XYZ) along with its corresponding surface normal information.

The proposed method begins by constructing a dual-latent diffusion model, which serves as an end-to-end conditional diffusion framework tailored for polyhedral point cloud generation. Subsequently, leveraging the generated polycube point clouds, we extract the underlying polycube structures 
which are then utilized to facilitate the final generation of all-hexahedral meshes
\begin{figure*}
    \centering
    \includegraphics[width=0.8\linewidth, trim=0.00cm 1.13cm 0.00cm 0.28cm, clip]{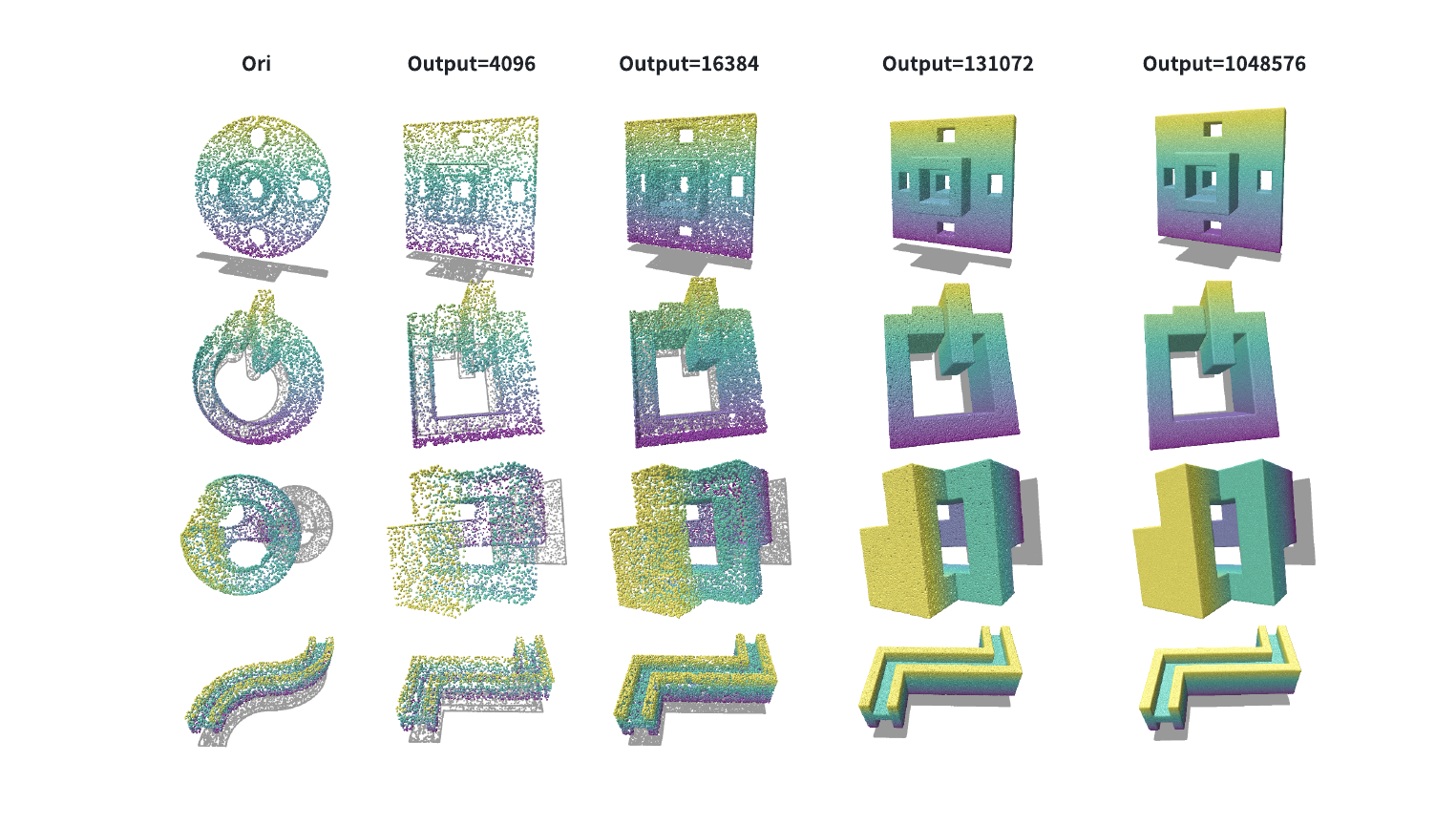}
    \caption{Results of polycube generation under a fixed input resolution of 4096 and incrementally increasing output polycube resolutions ranging from 4096 to 1,048,576.}
    \label{fig:fixed_input_4096}
\end{figure*}
\subsection{Dual-Latent diffusion model}
In contrast to DDPM-Polycube that relies on categorical one-hot labels for conditioning, our method shifts to leveraging the raw geometric point cloud $g$ as the conditioning signal. This guides the model to generate the corresponding polyhedral point cloud $x$ directly, effectively circumventing the need for explicit surface segmentation or predefined polycube templates. Furthermore, inspired by the efficiency of the two-stream block as demonstrated in Pointinfinity \cite{pointinfinity}, we introduce a dual-implicit conditional diffusion architecture. As illustrated in Fig. \ref{fig:arch}, this architecture not only transitions the denoiser backbone to the latent space but also maps the conditioning features into a low-dimensional latent manifold with a point cloud encoder, thereby enhancing both computational efficiency and representation power.

As shown in Fig.\ref{fig:arch},  given the data representations $(z^l, x^l)$, the architecture of the  two-stream block  first employs a read cross-attention block to encode the spatial information of the data into the latent space $\hat{z}^l$ as follows:$$\hat{z}^l = \text{CrossAtten}(z^l, x^l, x^l).$$ Subsequently, $H$ layers of Transformer self-attention are utilized to facilitate deep information fusion, yielding the updated latent state:$$z^{l+1} := \text{Transformer}(\hat{z}^l).$$
Finally, a write cross-attention block is applied to propagate the processed information from the latent space $z^{l+1}$ back into the data space $x^{l+1}$:$$x^{l+1} = \text{CrossAtten}(x^l, z^{l+1}, z^{l+1}).$$

 We define the operation of a two-stream block with the above three components  as:$$(x^{l+1}, z^{l+1}) = \text{TS}(x^l, z^l).$$

 As shown in Fig. \ref{fig:arch}, the Dual-Latent Diffusion Model, denoted by $\epsilon_{\theta}(x_t, e(g), t)$, is a parametric transformer-based architecture. Both the noise predictor $\epsilon_{\theta}$ and the condition encoder $e$ are constructed using the two-stream blocks. Specifically, the $l$-th layer of the condition encoder $e$ processes the conditional input $g^l$ and its associated latent state ${\tilde z}^l$ through a two-stream backbone, formulated as:$$(g^{l+1}, {  z}^{l+1}) =  \text{TS}(g^l, {  z}^l),$$ where $g^0=g$ and $z^0= z_{init}$ denote the initialized latent  embedding vectors.  The $l$-th layer of the  noise predictor $\epsilon_{\theta}$ processes the noisy input $x$ and its corresponding latent state $\tilde z^l$ via a similar two-stream backbone:
 
 $$(\tilde x^{l+1}, \tilde z^{l+1}) = \tilde{\text{TS}}(\tilde x^l, \tilde z^l),$$
where $\tilde x^0 = x_t$ and $\tilde z^0 = [ z^c, \tilde z_{init}, z_t={\rm embed}(t)]$. 
Here, $z^c$ denotes the output of the encoder, 
$\tilde z_{init}$ denotes the initialized latent embedding vector, 
and ${\rm embed}(t)$ denotes the embedding of the time step $t$.

The Dual-Latent Diffusion Model offers several significant improvements over DDPM-Polycube:

    
    

\noindent\textbf{Template-Free Conditional Generation:}
By utilizing raw geometric point clouds as the conditioning signal ($g$), our approach bypasses the need for explicit surface segmentation or predefined polycube templates.

\noindent\textbf{Linear Computational Scalability:}
The dual-latent architecture shifts the primary computational burden of condition encoding and noise prediction to a fixed-dimensional latent stream. By decoupling the complexity from dense spatial grids, the model’s computational overhead scales linearly with point resolution.

\noindent\textbf{Topological Feature Extraction \& Redundancy Filtering:}
Pol\-ycube structures are inherently characterized by planar regularity, which often results in significant informational redundancy in raw point cloud representations. Our framework effectively filters this noise by mapping the conditioning signal into a latent space. This transition allows the model to prioritize global topological consistency and key structural features over redundant local data, enabling the support of arbitrary input/output resolutions while maintaining structural integrity.
    
    

\begin{figure*}
    \centering
    \includegraphics[width=0.9\linewidth, trim=0.00cm 0.70cm 0.00cm 0.1cm, clip]{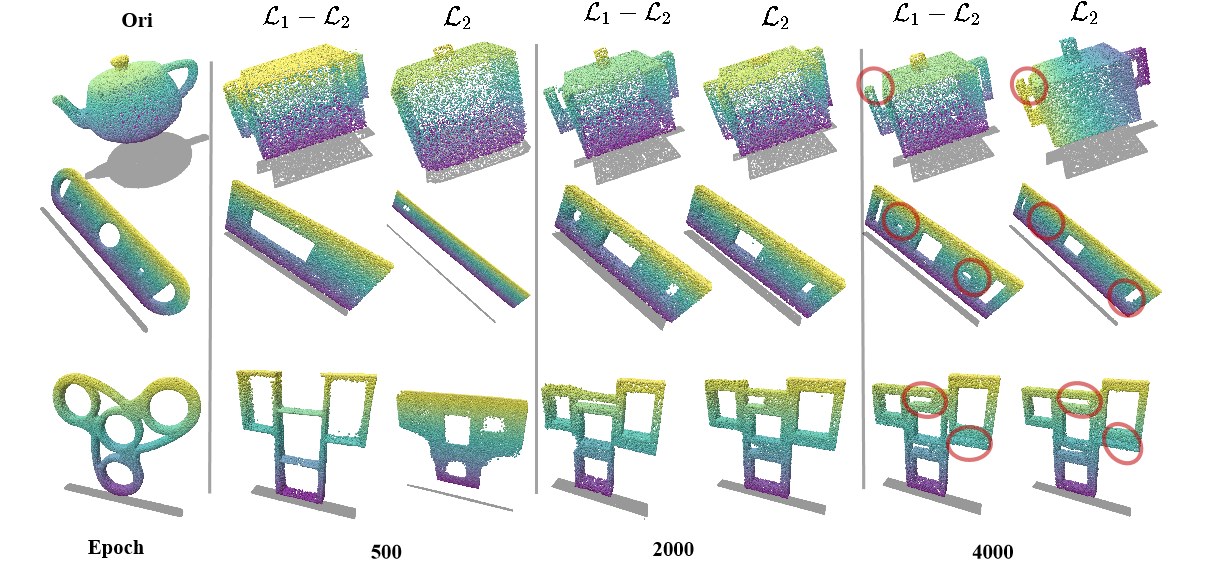}
    \caption{Inference outcomes comparison between dynamically weighted $\mathcal{L}_1$-$\mathcal{L}_2$ loss versus $\mathcal{L}_2$ loss throughout the training progression.}
    \label{fig:loss_comp}
\end{figure*}
 \subsection{Polycube Dataset}

We build a dataset for polycube mesh generation and analysis, starting from raw CAD triangle meshes. Our pipeline covers data collection, cleaning, polycube construction using multiple methods, robust deformation and geometry repair, and final registration and correspondence establishment. This dataset provides high-quality, structurally reliable samples for learning-based approaches and polycube-based hexahedral meshing.

We primarily collect meshes from the public ABC dataset \cite{ABCdataset} and conduct systematic topological and geometric quality checks. The meshes are categorized into multi-component shapes (i.e., containing multiple connected components), non-manifold meshes, open surfaces, meshes with foldovers, and meshes with degenerate elements. Only meshes with good geometric/topological quality and native polycube meshes are retained for the main dataset; the remaining cases are archived for difficult-case analysis.

We adopt two primary polycube construction methods. The first is deformation-based methods \cite{Fu16polycube}, which directly deform the input shape to obtain a polycube mesh. The second is label-based methods, where we first compute a valid label assignment on the original mesh \cite{expand_space,Evocube,CVTpolycube}, and then apply the robust polycube mapping method in \cite{Robust_quanti} to deform the labeled mesh into the polycube domain, producing the final polycube mesh. For each model, multiple candidate polycubes are generated using these methods, and the polycube with a foldover-free and low-distortion map is selected as the ground truth for the dataset.

For models requiring high-connectivity polycube structures \cite{Hexmesh_survey}, we first detect and handle these cases using \cite{expand_space}. If automatic processing fails, we use an interactive construction approach \cite{Interactive_cuboid} or discard the model when necessary.

To balance resolution and surface quality, we simplify and remesh the polycube surface, then establish the final correspondence between the original mesh and the polycube mesh. We have built an initial dataset of approximately 30K models, with plans to expand it further by incorporating additional shape repositories, such as Thingi10K \cite{Thingi10K}, to increase the diversity of polycube structures.

\subsection{Polycube-to-Hex Pipeline}

In the Polycube-to-Hex stage, we convert the polycube point cloud predicted by PolycubeNet into a hexahedral mesh via the following post-processing pipeline. Given an input triangular surface mesh $M$, we first sample an original point cloud $P_{\mathrm{ori}}$ on $M$ using Poisson-disk sampling \cite{sample_point}. Candidate points are drawn with respect to the surface-area measure and filtered by a minimum inter-point distance to obtain an approximately uniform distribution; we then enforce a fixed point count via random trimming and compensation sampling. For each sample, we store its incident face index and barycentric coordinates, yielding an exact and indexable correspondence between $P_{\mathrm{ori}}$ and $M$.

Conditioned on $P_{\mathrm{ori}}$, PolycubeNet predicts a polycube-structured point cloud $P_{\mathrm{poly}}$. We remove a small number of outliers using a lightweight local density filter to obtain $P_{\mathrm{poly}}^{\mathrm{clean}}$; due to the strong axis-aligned, piecewise-planar regularity of polycube outputs, this simple denoising is sufficient in practice. We then establish a cross-domain mapping via staged registration between $P_{\mathrm{ori}}$ and $P_{\mathrm{poly}}^{\mathrm{clean}}$: rigid alignment followed by non-rigid refinement with Coherent Point Drift (CPD) \cite{CPD}. The resulting continuous deformation field is discretized into a stable pointwise correspondence via nearest-neighbor projection. Combined with the stored barycentric coordinates, this yields a reliable mapping from $P_{\mathrm{poly}}^{\mathrm{clean}}$ to the input surface $M$.

Next, we recover an explicit polycube structure from $P_{\mathrm{poly}}^{\mathrm{clean}}$ via plane fitting and axis-aligned patch assembly, and generate a structured hexahedral mesh $H_1$ by regular partitioning in the polycube domain. To pull $H_1$ back to the input geometry, we follow a constrained optimization strategy inspired by prior hex-meshing pipelines \cite{Medial_xugang,HexOpt}. Keeping the hex connectivity fixed, we anchor boundary vertices of $H_1$ onto $M$ through the established mapping and jointly optimize surface conformity and interior element quality. Finally, we apply pillowing and smoothing for quality enhancement and boundary stabilization, following \cite{Robust_quanti}.

\begin{figure*}
    \centering
    \includegraphics[width=0.9\linewidth, trim=0.00cm 0.03cm 0.00cm 0.28cm, clip]{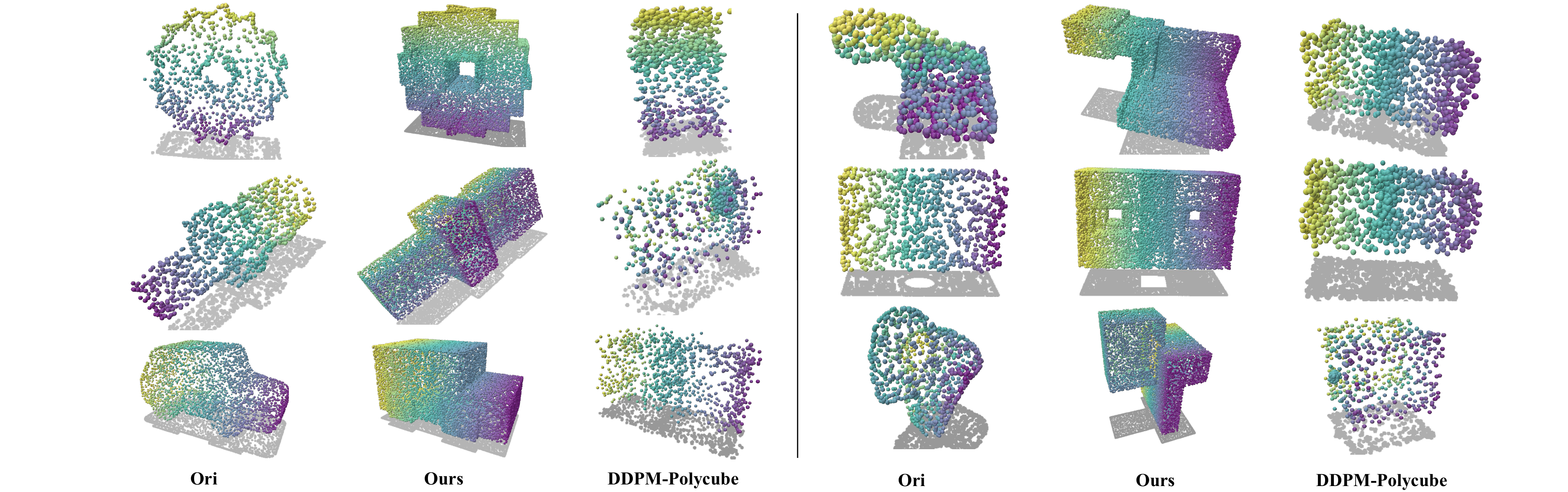}
    \caption{Results from PolycubeNet(our) against DDPM-Polycube, using input geometry at a resolution of 1024.}
    \label{fig:compare_with_ddpm}
\end{figure*}
\section{Implementation details}

\subsection{Architecture Details}

\textbf{Denoiser.}  The core denoiser consists of $L = 6$ two-stream blocks. Each of these blocks integrates $H = 4$ transformer layers. We utilize a hidden dimension of $d_{model} = 256$ across all layers, with multi-head attention configured with $8$. The model operates on $N_{latent} = 256$ tokens, each with a dimensionality of $d_{latent} = 256$. 
 
\noindent\textbf{Conditioning Encoder.}  For the conditional input, we employ a dedicated encoder comprising $6$ two-stream blocks and $6$ transformer layers in each block with $8$ heads. This encoder transforms raw inputs into a sequence of $64$ latent tokens ($d = 256$).

\subsection{Training Loss}

\textbf{Training  Parameters.} The model is trained with a batch size of $32$. Both the resolution of the training data $x$ and the geometry condition $g$ are set to $4096$, that is, $N=M=4096$. We use the AdamW optimizer with an initial learning rate of $1 \times 10^{-4}$.   A cosine learning rate decay schedule is applied throughout training to ensure stable convergence. The max time step $T$ is set to 1024.

\noindent\textbf{Training Loss.} In the training of diffusion models, the primary objective function typically adopts the $\mathcal L_2$ loss based on Gaussian error, formulated as follows:
$$\mathcal{L}_{\text{2}} = \mathbb{E}_{t, \mathbf{x}_0, g, \boldsymbol{\epsilon}} \left[ \| \boldsymbol{\epsilon} - \boldsymbol{\epsilon}_\theta( \sqrt{\bar{\alpha}_t} \, \mathbf{x}_0 + \sqrt{1 - \bar{\alpha}_t} \, \boldsymbol{\epsilon},g, t ) \|^2 \right].$$

We introduce a dynamically weighted $\mathcal{L}_1$–$\mathcal{L}_2$ hybrid loss. This loss function incorporates a stochastic weighting mechanism during each training iteration, integrating the $\mathcal{L}_1$ loss (mean absolute error) with the $\mathcal{L}_2$ loss. The formulation is defined as:$$w \sim \mathcal{U}[0.4, 0.8), \quad \mathcal{L}_{\text{hybrid}} = w\cdot \mathcal{L}_{\text{2}} + (1 - w) \cdot \mathcal{L}_{\text{1}}.$$
The weight is uniformly sampled from the interval [0.4, 0.8) to maintain a dynamic balance between the loss components during training. Numerical experiments demonstrate that the dynamically weighted loss converges faster than the $\mathcal{L}_2$ loss.

\subsection{Point Cloud Post-Processing: Outlier Removal}

To ensure the precision of the generated model, we implement a two-phase outlier filtering procedure to refine the point cloud $P = \{p_i \in \mathbb{R}^3\}_{i=1}^N$ after generation:

\begin{itemize}
    \item \textbf{Phase I: Local Connectivity Detection} \\
    We define an $L_1$-neighborhood decision function to identify isolated noise:
     $$
        \mathcal{N}(p_i) = \{ p_j \in P \mid \| p_i - p_j \|_1 < \tau \},
   $$
    where $\tau$ is the distance threshold. If $|\mathcal{N}(p_i)| = 0$, $p_i$ is classified as an absolute outlier and removed. 

    \item \textbf{Phase II: Neighborhood Density Filtering} \\
    For the filtered set $P'$, we further calculate the minimum proximity distance for each point:
     
       $$ d_{\min}(p_i) = \min_{p_j \in \mathcal{N}(p_i)} \| p_i - p_j \|_1.$$
    
    The points are then sorted by $d_{\min}$ in ascending order. The points corresponding to the $K$ largest values  are pruned. This step effectively eliminates relatively isolated points residing within locally sparse regions.
\end{itemize}

\section{Experiments}

\subsection{Main results of PolycubeNet for polycube generation. }

\textbf{Test-Time Resolution Scaling.} Fig.~\ref{fig:fixed_output_resolution} illustrates a comparison under fixed input conditions, where the output resolution is progressively increased from $4,096$ to $1,048,576$. PolycubeNet consistently generates high-quality polycubes across this range. Conversely, Fig.~\ref{fig:fixed_input_4096} evaluates the performance with a fixed output resolution of $131,072$ while varying the input resolution from $128$ to $16,384$.  We observe that once the input resolution reaches 1,024, which is sufficient to capture the necessary geometric information, the quality of the generated polycube becomes comparable to that achieved with a resolution of 16,384. Although the model was trained with both input and output resolutions fixed at $4,096$, it supports variable resolutions during inference, maintaining the ability to produce high-quality polycube point clouds.

\begin{figure}
    \centering
    \includegraphics[width=0.76\linewidth, trim=0.00cm 0.3cm 0.00cm 0.09cm, clip]{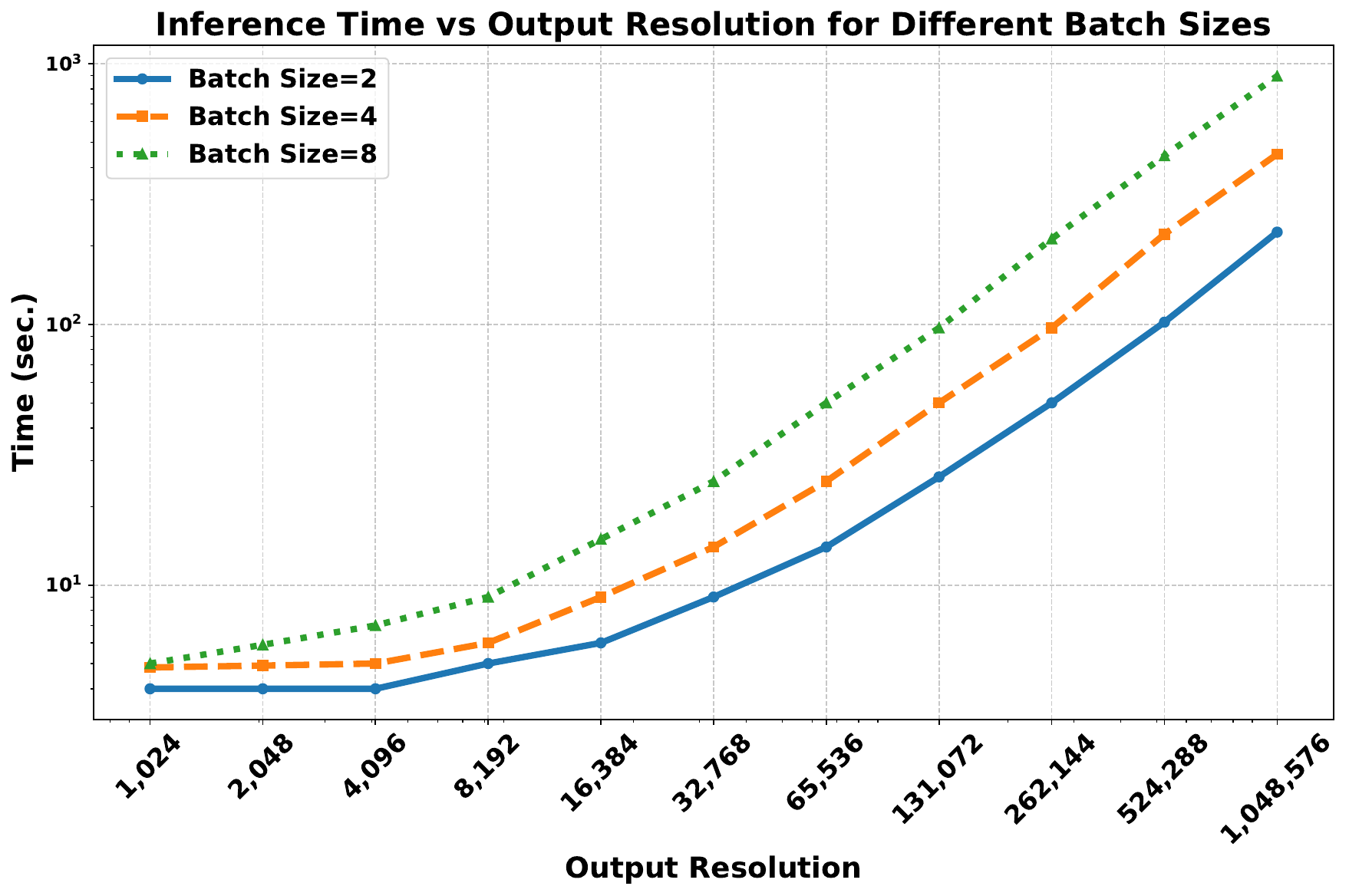}
    \caption{The curves depict the relationship between output resolution and inference time for different batch sizes. All experiments use DDIM \cite{DDIM} sampling with a stride of 4 between timesteps.}
    \label{fig:time_plot}
\end{figure}

\noindent\textbf{Linear Time Complexity.} Fig.~\ref{fig:time_plot} presents the variation of inference time of PolycubeNet as the generated resolution increases from 1024 to 1,048,576. The results demonstrate that the inference time complexity of our model increases linearly with the improvement in resolution.

\noindent\textbf{Comparisons of Training losses.}  Fig.~\ref{fig:loss_comp} compares the inference results of the weighted dynamic $\mathcal{L}_1$-$\mathcal{L}_2$  loss and the $\mathcal{L}_2$-only loss during the training process. The results demonstrate that the weighted dynamic $\mathcal{L}_1$-$\mathcal{L}_2$  loss achieves  faster convergence rate than the $\mathcal{L}_2$ loss. This advantage is manifested in the superior capture of geometric topology during the early stages of training, as well as the enhanced representation of fine details in the later stages.

\noindent\textbf{Comparisons with DDPM-Polycube.} Fig.~\ref{fig:compare_with_ddpm} demonstrates the generation results of PolycubeNet and DDPM-Polycube on several models. The results indicate that the polycubes generated by our model significantly outperform those produced by DDPM-Polycube in terms of quality, geometric and topological structure, as well as detail preservation.

\subsection{Comparison with Classical Polycube Methods}

To evaluate our method against classical polycube pipelines, we compare polycube structural correctness, coverage of failure modes, and hex-mesh quality. Unless otherwise stated, we use \cite{Evocube} Evocube as the primary baseline, whose workflow typically relies on surface segmentation, axis-label assignment, and subsequent continuous optimization.

\begin{figure}
    \centering
    \includegraphics[width=0.8\linewidth, trim=0.00cm 0.93cm 0.00cm 0.91cm, clip]{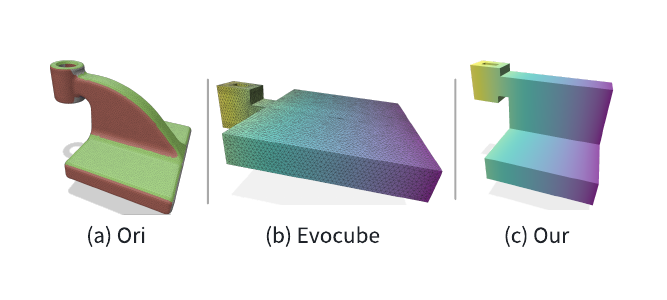}
    \caption{Polycube comparison on a slanted feature. (a) Axis-label assignment from Evocube. (b) Polycube result of Evocube. (c) Our predicted polycube.}
    \label{fig:compare_tra}
\end{figure}

\noindent\textbf{Structural artifacts and global bias.} As noted in \cite{Hexmesh_survey}, classical polycube methods can be steered by local cues toward an incorrect global structure, which is often difficult to recover from during optimization. Challenging patterns include small but topologically critical features (e.g., tunnels and handles), tiny protrusions, and long yet slightly slanted creases that misdirect the axis-alignment drive, leading to inconsistent solutions known as structural artifacts \cite{Hexmesh_survey}. Figure~\ref{fig:compare_tra} shows a representative example: (a) visualizes the principal-axis labels produced by Evocube on the input surface, and (b) shows the resulting deformation where the ramp collapses into a single plane, causing structural distortion. In contrast, our result in (c) preserves a polycube structure consistent with the overall shape.


\begin{figure}[!t]
    \centering
    \includegraphics[width=0.54\linewidth, trim=0.90cm 0.89cm 0.60cm 0.93cm, clip]{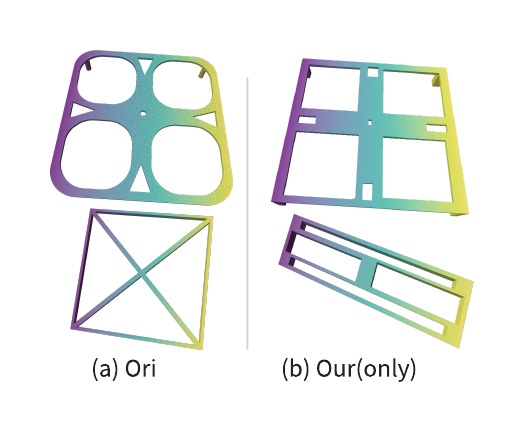}
    \caption{Classical pipelines fail; ours remains valid.}
    \label{fig:just_our}
\end{figure}


\noindent\textbf{Solvability under polycube-map conflicts.} We also observe infeasible cases induced by polycube-map conflicts, where axis-alignment and connectivity constraints across local regions are incompatible, making heuristic/local optimization pipelines fail to find a feasible solution. Figure~\ref{fig:just_our} presents two examples: one contains a "triangular" inner boundary, and the other exhibits conflicts between an outer "square" silhouette and an inner "X" structure. Classical methods often fail to produce a topologically consistent polycube under their validity assumptions and local fixing strategies, whereas our approach can still directly generate a globally consistent polycube output, expanding the set of shapes that can be robustly mapped to a polycube domain.

\noindent\textbf{Efficiency.} Classical pipelines usually involve multiple rounds of segmentation/label search and continuous optimization, with runtime increasing notably with shape complexity. Representative methods such as \cite{Polycut,Evocube,close_form_polycube} often take from a few minutes to tens of minutes in our experiments. In contrast, once trained, our diffusion model generates the polycube point cloud within seconds at inference, which is favorable for batch processing and interactive iterations.

\begin{figure*}
    \centering
    \includegraphics[width=1\linewidth, trim=0.00cm 0.9cm 0.00cm 1.0cm, clip]{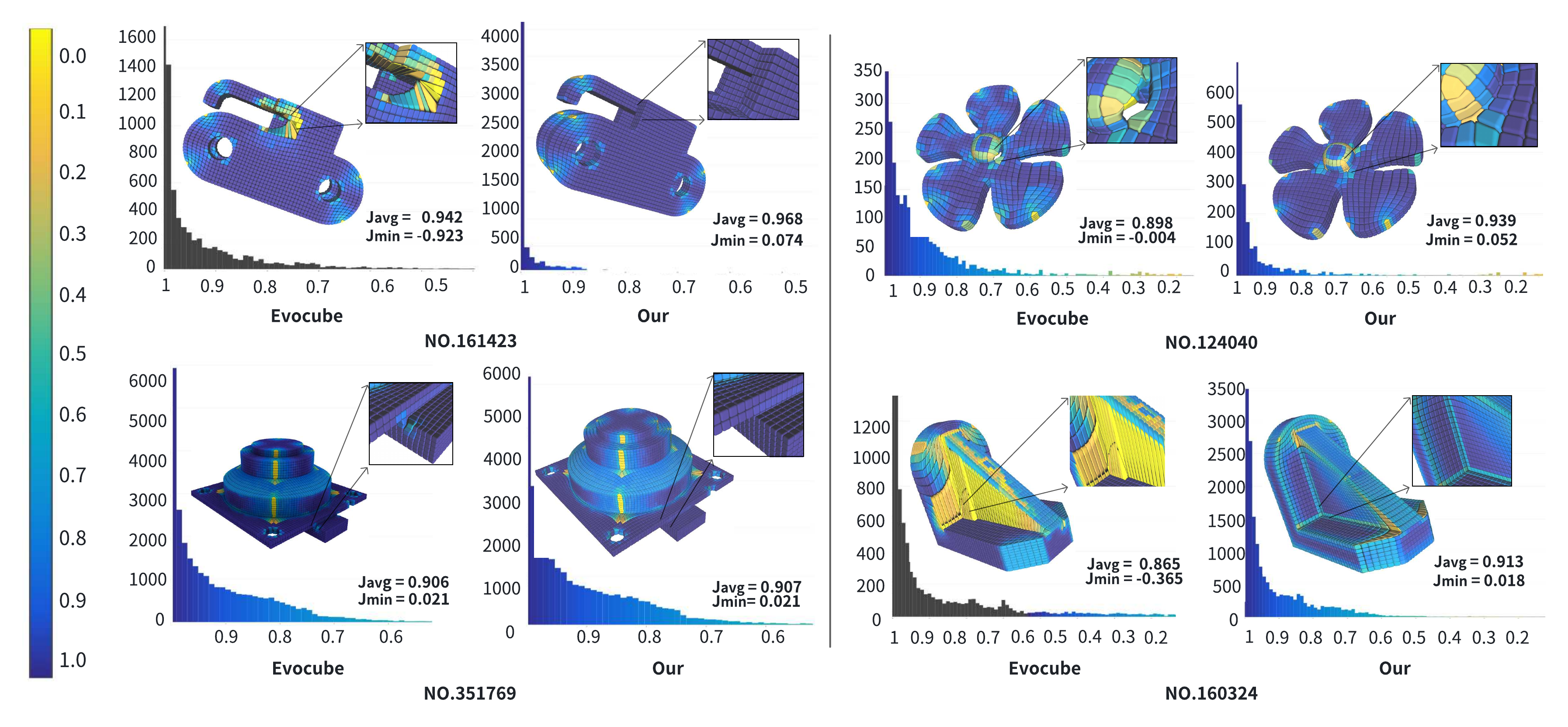}
    \caption{Hex-mesh comparison. Hex meshes generated from our polycube outputs and from Evocube, visualized with HexaLab. We report scaled Jacobian statistics ($J_{\mathrm{avg}}$, $J_{\min}$).}
    \label{fig:compare_hex}
\end{figure*}

\noindent\textbf{Hex-mesh quality.} Figure~\ref{fig:compare_hex} compares the hexahedral meshes produced by our method and Evocube, with consistent visualization using \cite{HexaLab} HexaLab. We report the average and minimum scaled Jacobian, $J_{\mathrm{avg}}$ and $J_{\min}$. Overall, our method achieves better $J_{\mathrm{avg}}/J_{\min}$ and avoids structural failures of the baseline: e.g., Evocube exhibits a topological issue on No.12404, and severe distortion induced by structural artifacts on No.160324. For the zoomed regions of No.161423 and No.351769, the baseline produces a 4-connectivity corner, violating its typical validity assumption (corners are restricted to 3-connectivity), thus requiring engineering constraints or fixes and introducing mapping distortion that degrades hex quality. While \cite{expand_space} extends the validity space for certain high-connectivity cases, it still relies on local patching and engineering rules and does not cover the full polycube space. In contrast, our learned global structural prior improves robustness over a broader set of shapes and yields more consistent quality gains.

\noindent\textbf{Positioning.} Our goal is not to replace optimization-based polycube pipelines: classical methods remain advantageous in interpretability, determinism, and enforcing specific engineering constraints, and Evocube demonstrates strong practical performance. We instead provide a generative framework for AI for Hex that learns and outputs globally consistent structural priors, complementing classical pipelines and enabling future hybrid approaches that combine learning with optimization.
\section{limitations}



\noindent\textbf{Narrow gaps and thin structures.} Similar to many point-cloud pipelines, our method operates on unstructured samples without explicit mesh connectivity, which can make extremely narrow slits hard to disambiguate. Points from opposing sides may become spatially indistinguishable, leading to incorrect separation and degraded downstream polycube/hex construction. A natural direction is to augment points with connectivity cues (e.g., neighborhood graphs) and incorporate normal-aware constraints to better preserve thin-feature separation.

\noindent\textbf{Training distribution coverage.} Our training data are primarily CAD-like shapes from the ABC dataset, with limited coverage of artistic models and other non-CAD distributions. Expanding the dataset with additional repositories and curated categories is expected to improve robustness under broader geometry styles and reduce distribution shift.

\section{Conclusions}

In summary, PolycubeNet presents an end-to-end, diffusion-based framework for polycube generation conditioned on input geometry, together with a dual-latent Transformer design that decouples the dominant computation from sampling resolution. This design confines expensive denoising and conditioning operations to a fixed-capacity latent space, enabling efficient generation while preserving global structural consistency. Unlike classical polycube pipelines that depend on surface labeling and heuristic optimization, we follow a structure-first strategy that directly synthesizes polycube representations and then establishes stable correspondences via staged registration, which reduces failure modes caused by brittle intermediate steps. Building on the generated polycube, we further provide a complete polycube-to-hex pipeline that recovers explicit polycube structure, performs regular partitioning, and pulls the structured mesh back to the input surface through constrained optimization, followed by standard quality enhancement. Across our experiments, PolycubeNet demonstrates improved robustness and practical efficiency on diverse CAD geometries, and yields high-quality polycube structures that support reliable downstream hexahedral meshing.

To mitigate the scarcity of real polycube supervision, we build and will release the first paired dataset of CAD meshes and their corresponding polycube meshes, together with the core code. More broadly, our results suggest that learning structured intermediate representations with generative modeling can provide a viable path for tackling mesh problems governed by discrete, non-local constraints, where directly predicting final elements is often unstable. Looking forward, we plan to integrate diffusion-based structure prediction with classical optimization to combine the flexibility of learning and the controllability of traditional pipelines, potentially enabling more reliable refinement under application-specific constraints. We also aim to extend this “learned intermediate representation” paradigm to other mesh tasks involving discrete structural or topological predictions, while broadening data coverage to improve generalization across diverse geometry styles.



\begin{thebibliography}{1}
\bibliographystyle{IEEEtran}

\bibitem{Hexmesh_survey}
N. Pietroni, M. Campen, A. Sheffer, G. Cherchi, D. Bommes, X. Gao, R. Scateni, F. Ledoux, J. Remacle, and M. Livesu, ``Hex-Mesh Generation and Processing: A Survey,'' \textit{ACM Trans. Graph.}, vol. 42, no. 2, pp. 1--44, 2023.

\bibitem{NASA_report}
J. P. Slotnick, A. Khodadoust, J. J. Alonso, D. L. Darmofal, W. D. Gropp, E. A. Lurie, and D. J. Mavriplis, \textit{CFD Vision 2030 Study: A Path to Revolutionary Computational Aerosciences}, NASA/CR-2014-218178, NASA, 2014.

\bibitem{HolyGrailHex}
T. D. Blacker, ``Meeting the Challenge for Automated Conformal Hexahedral Meshing,'' in \textit{Proc. 9th Int. Meshing Roundtable}, 2000, pp. 11--20.

\bibitem{polycubemap04}
M. Tarini, K. Hormann, P. Cignoni, and C. Montani, ``Polycube-maps,'' \textit{ACM Trans. Graph.}, vol. 23, no. 3, pp. 853--860, 2004.

\bibitem{deform_polycube2011}
J. Gregson, A. Sheffer, and E. Zhang, ``All-Hex Mesh Generation via Volumetric PolyCube Deformation,'' \textit{Comput. Graph. Forum}, 2011.

\bibitem{Polycut}
M. Livesu, N. Vining, A. Sheffer, J. Gregson, and R. Scateni, ``PolyCut: Monotone Graph-Cuts for PolyCube Base-Complex Construction,'' \textit{ACM Trans. Graph.}, vol. 32, no. 6, pp. 1--12, 2013.

\bibitem{HexEx}
M. Lyon, D. Bommes, and L. Kobbelt, ``HexEx: robust hexahedral mesh extraction,'' \textit{ACM Trans. Graph.}, vol. 35, no. 4, 2016.

\bibitem{15fix_normal_polycube}
D. Sokolov and N. Ray, ``Fixing Normal Constraints for Generation of Polycubes,'' 2015.

\bibitem{close_form_polycube}
X. Fang, W. Xu, H. Bao, and J. Huang, ``All-hex Meshing using Closed-form Induced Polycube,'' \textit{ACM Trans. Graph.}, vol. 35, no. 4, pp. 1--9, 2016.

\bibitem{DL-Polycube}
Y. Yu, Y. Fang, H. Tong, and Y. J. Zhang, ``DL-Polycube: deep learning enhanced generalized polycube method for high-quality hexahedral mesh generation and volumetric spline construction,'' \textit{Eng. Comput.}, pp. 1--24, 2025.

\bibitem{DDPM-polycube}
Y. Yu, Y. Fang, H. Tong, J. Liu, and Y. Zhang, ``DDPM-Polycube: a denoising diffusion probabilistic model for polycube-based hexahedral mesh generation and volumetric spline construction,'' \textit{Eng. Comput.}, vol. 42, 2025.

\bibitem{DDPM}
J. Ho, A. Jain, and P. Abbeel, ``Denoising diffusion probabilistic models,'' \textit{Adv. Neural Inf. Process. Syst.}, vol. 33, pp. 6840--6851, 2020.

\bibitem{pointinfinity}
Z. Huang, J. Johnson, S. Debnath, J. M. Rehg, and C.-Y. Wu, ``PointInfinity: Resolution-Invariant Point Diffusion Models,'' in \textit{Proc. IEEE/CVF Conf. Comput. Vis. Pattern Recognit.}, 2024, pp. 10050--10060.

\bibitem{L1polycube14}
J. Huang, T. Jiang, Z. Shi, Y. Tong, H. Bao, and M. Desbrun, ``l1-based Construction of Polycube Maps from Complex Shapes,'' \textit{ACM Trans. Graph.}, vol. 33, no. 3, pp. 1--11, 2014.

\bibitem{L1polycube19}
L. Chen, G. Xu, S. Wang, Z. Shi, and J. Huang, ``Constructing Volumetric Parameterization based on Directed Graph Simplification of $\ell_1$ Polycube Structure from Complex Shapes,'' \textit{Comput. Methods Appl. Mech. Engrg.}, vol. 351, pp. 422--440, 2019.

\bibitem{Evocube}
C. Dumery, F. Protais, S. Mestrallet, C. Bourcier, and F. Ledoux, ``Evocube: a Genetic Labeling Framework for Polycube-Maps,'' \textit{Comput. Graph. Forum}, vol. 41, no. 6, pp. 467--479, 2022.

\bibitem{fu19polycube}
Y. Yang, X.-M. Fu, and L. Liu, ``Computing Surface PolyCube-Maps by Constrained Voxelization,'' in \textit{Comput. Graph. Forum}, vol. 38, no. 7, pp. 299--309, 2019.

\bibitem{Intrinsic_polycube}
M. Mandad, R. Chen, D. Bommes, and M. Campen, ``Intrinsic mixed-integer polycubes for hexahedral meshing,'' \textit{Comput. Aided Geom. Des.}, vol. 94, p. 102078, 2022.

\bibitem{CVTpolycube}
K. Hu, Y. Zhang, and T. Liao, ``Surface segmentation for polycube construction based on generalized centroidal Voronoi tessellation,'' \textit{Comput. Methods Appl. Mech. Engrg.}, vol. 316, pp. 280--296, 2017.

\bibitem{Interactive_cuboid}
L. Li, P. Zhang, D. Smirnov, S. M. Abulnaga, and J. Solomon, ``Interactive All-Hex Meshing via Cuboid Decomposition,'' \textit{ACM Trans. Graph.}, vol. 40, no. 6, pp. 1--17, 2021.

\bibitem{expand_space}
L. He, N. Lei, Z. Wang, C. Wang, X. Zheng, and Z. Luo, ``High-connectivity polycube-maps: Solvable space expansion through validity-augmented topological conditions,'' \textit{Comput.-Aided Des.}, vol. 191, p. 103972, 2026.

\bibitem{MeshWalker}
A. Lahav and A. Tal, ``Meshwalker: Deep mesh understanding by random walks,'' \textit{ACM Trans. Graph.}, vol. 39, no. 6, pp. 1--13, 2020.

\bibitem{MeshCNN}
R. Hanocka, A. Hertz, N. Fish, R. Giryes, S. Fleishman, and D. Cohen-Or, ``MeshCNN: a network with an edge,'' \textit{ACM Trans. Graph.}, vol. 38, no. 4, pp. 1--12, 2019.

\bibitem{DiffusionNet}
N. Sharp, S. Attaiki, K. Crane, and M. Ovsjanikov, ``DiffusionNet: Discretization agnostic learning on surfaces,'' \textit{ACM Trans. Graph.}, vol. 41, no. 3, pp. 1--16, 2022.

\bibitem{Point2Mesh}
R. Hanocka, G. Metzer, R. Giryes, and D. Cohen-Or, ``Point2Mesh: A self-prior for deformable meshes,'' arXiv:2005.11084, 2020.

\bibitem{PointTriNet}
N. Sharp and M. Ovsjanikov, ``PointTriNet: Learned triangulation of 3D point sets,'' in \textit{Proc. Eur. Conf. Comput. Vis.}, 2020, pp. 762--778.

\bibitem{RL_quad}
H. Tong, K. Qian, E. Halilaj, and Y. J. Zhang, ``SRL-assisted AFM: Generating planar unstructured quadrilateral meshes with supervised and reinforcement learning-assisted advancing front method,'' \textit{J. Comput. Sci.}, vol. 72, p. 102109, 2023.

\bibitem{ho2020denoising}
J. Ho, A. Jain, and P. Abbeel, ``Denoising diffusion probabilistic models,'' \textit{Adv. Neural Inf. Process. Syst.}, vol. 33, pp. 6840--6851, 2020.

\bibitem{rombach2022high}
R. Rombach, A. Blattmann, D. Lorenz, P. Esser, and B. Ommer, ``High-resolution image synthesis with latent diffusion models,'' in \textit{Proc. IEEE/CVF Conf. Comput. Vis. Pattern Recognit.}, 2022, pp. 10684--10695.

\bibitem{saharia2022photorealistic}
C. Saharia, W. Chan, S. Saxena, L. Li, J. Whang, E. L. Denton, K. Ghasemipour, R. Gontijo Lopes, B. Karagol Ayan, T. Salimans, \textit{et al.}, ``Photorealistic text-to-image diffusion models with deep language understanding,'' \textit{Adv. Neural Inf. Process. Syst.}, vol. 35, pp. 36479--36494, 2022.

\bibitem{blattmann2023stable}
A. Blattmann, T. Dockhorn, S. Kulal, D. Mendelevitch, M. Kilian, D. Lorenz, Y. Levi, Z. English, V. Voleti, A. Letts, \textit{et al.}, ``Stable video diffusion: Scaling latent video diffusion models to large datasets,'' arXiv:2311.15127, 2023.

\bibitem{wan2025wan}
Team Wan, A. Wang, B. Ai, B. Wen, C. Mao, C.-W. Xie, D. Chen, F. Yu, H. Zhao, J. Yang, \textit{et al.}, ``Wan: Open and advanced large-scale video generative models,'' arXiv:2503.20314, 2025.

\bibitem{liu2023zero}
R. Liu, R. Wu, B. Van Hoorick, P. Tokmakov, S. Zakharov, and C. Vondrick, ``Zero-1-to-3: Zero-shot one image to 3D object,'' in \textit{Proc. IEEE/CVF Int. Conf. Comput. Vis.}, 2023, pp. 9298--9309.

\bibitem{tang2024mvdiffusion++}
S. Tang, J. Chen, D. Wang, C. Tang, F. Zhang, Y. Fan, V. Chandra, Y. Furukawa, and R. Ranjan, ``MVDiffusion++: A dense high-resolution multi-view diffusion model for single or sparse-view 3D object reconstruction,'' in \textit{Proc. Eur. Conf. Comput. Vis.}, 2024, pp. 175--191.

\bibitem{long2024wonder3d}
X. Long, Y.-C. Guo, C. Lin, Y. Liu, Z. Dou, L. Liu, Y. Ma, S.-H. Zhang, M. Habermann, C. Theobalt, \textit{et al.}, ``Wonder3D: Single image to 3D using cross-domain diffusion,'' in \textit{Proc. IEEE/CVF Conf. Comput. Vis. Pattern Recognit.}, 2024, pp. 9970--9980.

\bibitem{PVD}
S. Luo and W. Hu, ``Diffusion probabilistic models for 3D point cloud generation,'' in \textit{Proc. IEEE/CVF Conf. Comput. Vis. Pattern Recognit.}, 2021, pp. 2837--2845.

\bibitem{Point-E}
A. Nichol, H. Jun, P. Dhariwal, P. Mishkin, and M. Chen, ``Point-E: A system for generating 3D point clouds from complex prompts,'' arXiv:2212.08751, 2022.

\bibitem{ABCdataset}
S. Koch, A. Matveev, Z. Jiang, F. Williams, A. Artemov, E. Burnaev, M. Alexa, D. Zorin, and D. Panozzo, ``ABC: A Big CAD Model Dataset for Geometric Deep Learning,'' in \textit{Proc. IEEE/CVF Conf. Comput. Vis. Pattern Recognit.}, 2019.

\bibitem{Fu16polycube}
C.-Y. Bai, Y. Liu, and X.-M. Fu, ``Efficient Volumetric PolyCube-Map Construction,'' \textit{Comput. Graph. Forum}, 2016.

\bibitem{Robust_quanti}
F. Protais, M. Reberol, N. Ray, E. Corman, F. Ledoux, and D. Sokolov, ``Robust Quantization for Polycube Maps,'' \textit{Comput.-Aided Des.}, vol. 150, p. 103321, 2022.

\bibitem{Thingi10K}
Q. Zhou and A. Jacobson, ``Thingi10K: A Dataset of 10,000 3D-Printing Models,'' \textit{CoRR}, vol. abs/1605.04797, 2016.

\bibitem{sample_point}
C. Yuksel, ``Sample Elimination for Generating Poisson Disk Sample Sets,'' \textit{Comput. Graph. Forum}, vol. 34, no. 2, pp. 25--32, 2015.

\bibitem{CPD}
A. Myronenko and X. B. Song, ``Point Set Registration: Coherent Point Drift,'' \textit{IEEE Trans. Pattern Anal. Mach. Intell.}, vol. 32, pp. 2262--2275, 2010.

\bibitem{Medial_xugang}
S. Zhang, G. Xu, H. Wu, R. Gu, L. Qi, and Y. Pang, ``Medial hex-meshing: high-quality all-hexahedral mesh generation based on medial mesh,'' \textit{Eng. Comput.}, vol. 40, no. 4, pp. 2537--2557, 2024.

\bibitem{HexOpt}
H. Tong and Y. J. Zhang, ``Fast and Robust Hexahedral Mesh Optimization via Augmented Lagrangian, L-BFGS, and Line Search,'' \textit{CoRR}, vol. abs/2410.11656, 2024.

\bibitem{DDIM}
J. Song, C. Meng, and S. Ermon, ``Denoising Diffusion Implicit Models,'' arXiv:2010.02502, 2020.

\bibitem{HexaLab}
M. Bracci, M. Tarini, N. Pietroni, M. Livesu, and P. Cignoni, ``HexaLab.net: An online viewer for hexahedral meshes,'' \textit{Comput.-Aided Des.}, vol. 110, pp. 24--36, 2019.

\clearpage

\end{thebibliography}

%
%
%
%


\end{document}